\documentclass[aps, pra, twocolumn,10pt, superscriptaddress, showpacs]{revtex4-1}

\usepackage{epsfig}
\usepackage{amsmath}
\usepackage{amssymb}

\begin{document}

\author{Jacek Gruca}
\affiliation{Instytut Fizyki Teoretycznej i Astrofizyki, Uniwersytet Gda\'nski, 80-952 Gda\'nsk, Poland}

\author{Wies{\l}aw Laskowski}
\email{wieslaw.laskowski@univ.gda.pl}
\affiliation{Instytut Fizyki Teoretycznej i Astrofizyki, Uniwersytet Gda\'nski, 80-952 Gda\'nsk, Poland}

\author{Marek \.Zukowski}
\affiliation{Instytut Fizyki Teoretycznej i Astrofizyki, Uniwersytet Gda\'nski, 80-952 Gda\'nsk, Poland}
\affiliation{University of Science and Technology of China, Hefei, Anhui, China}

\title{Nonclassicality of pure two-qutrit entangled states}


\begin{abstract}
We report an exhaustive numerical analysis of violations of local realism  by two qutrits in all possible pure entangled states. In Bell type experiments we allow any pairs of local unitary U(3) transformations to define the measurement bases.  Surprisingly, Schmidt rank-2 states, resembling pairs of maximally entangled qubits, lead to the most noise-robust violations of local realism. The phenomenon seems to be even more pronounced for four and five dimensional systems, for which we tested a few interesting examples.
\end{abstract}

\maketitle

\section{Introduction}
John Bell \cite{BELL} has shown that  local realistic models are impossible for  quantum mechanics of two  qubits. After some years researchers started to
ask questions about the Bell theorem for more complicated systems (compare the classic review in which such problems are not covered, \cite{CLAUSER}).   First results concerning   possible extensions to  entangled states of pairs of $d$-state systems, 
with $d \geq 3$, arrived  in early eighties \cite{MERMIN}. 
The early research was confined to Stern-Gerlach
type measurements (which are parametrized by only two numbers, defining the direction of the measurement axis). Some authors speculated that correspondence principle suggests that non-classicality should diminish with growing $d$.   
However, on one hand in the meantime we witnessed a surprise in the form of the GHZ theorem \cite{GHZ}: for three or more qubits the conflict 
between local realism and quantum mechanics is much sharper than for two. Thus, with the increase of the dimensionality of the full system, non-classicality of quantum correlations can grow.
On the other hand,  Peres and Gisin \cite{PERES} considered  dichotomic observables applied to maximally entangled pairs of qu$d$its, and showed that  violations of the CHSH inequalities robustly survive in the limit of $d \to \infty$.

The concept of multiport interferometers, first  discussed in the context of quantum entanglement  by Klyshko \cite{KLYSHKO}, gave a hope for operational realizations of unitary transformations much richer than those linking measurement Stern-Gerlach bases for higher spins.
Proposals of Bell experiments with the multiports were presented in \cite{CONTROL, multiport}. Multiport interferometers were shown to be capable to reproduce all finite  dimensional unitary
transformations \cite{RECK}, i.e., one can have access to the full U($d$) group. This  lead to the discovery that two maximally entangled  qu$d$its violate local realism more strongly than qubits, and that this violation grows with $d$, see \cite{KASZLIKOWSKI}. The strength of the non-classicality was measured via a ``white'' noise resistance. One takes 
a family of mixed states $
%\begin{equation}
\varrho= v \varrho_{state} + (1-v)\varrho_{noise},
%\end{equation}
$
where $\varrho_{state}$ represents the state under consideration, and  $\varrho_{noise}=\frac{1}{d^2}\openone$ is the maximally mixed state. The lower is the threshold value $v=v_{crit}$ beyond which state does not violate local realism in a given Bell experiment, the bigger is the noise admixture $1 -v_{crit}$, which is needed to erase the non-classicality of the state in the experiment.
Note that $v_{crit}$ is a natural generalization of the ``factor of violation'' of a Bell inequality. E.g. for the CHSH inequality, if $\hat B$ is the Bell operator, its local realistic bound can be violated maximally by a factor of $v_{crit}^{-1}$. The same can be said about CGLMP inequalities \cite{CGLMP}. The results were obtained via a numerical analysis, which is a prototype  of the one which will be presented.  The values were confirmed by Collins et al. \cite{CGLMP} who derived a set of tight correlation Bell (CGLMP) inequalities specific for qu$d$it measurements. This tool helped to discover a strange property of two qu$d$it states. Maximally entangled states do not violate the CGLMP inequalities for qutrits maximally. The Schmidt decompositions of the optimal states does have all amplitudes of the same modulus --- one of them is smaller \cite{ACIN}. The results were generalized further by Chen at al. \cite{CHEN}. With a new version of a qu$d$it Bell inequality, the optimality of non-maximally entangled states was shown in arbitrary dimension, \cite{ZOHREN}.  Note however, that in  the  papers a specific set of local observables was used, defined by a set of local phase shifts (in the  Schmidt decomposition basis for the state), followed by an ``unbiased'' multiport beamsplitters (class M1 in Fig. 1), and finally detectors. Thus, non-classicality was not fully mapped.

Here we report an exhaustive numerical analysis of the two-qutrit Bell experiment involving two settings per observer. The observables are absolutely general, all possible ones are studied (that is the unitary transformations defining the measurement bases are forming the U(3) group). Such an analysis allows us to map the strength of violations of local realism (as measured by noise resistance) for all possible pure two-qutrit states. In this way we discover that Schmidt rank-2 states give rise to highest violations. This phenomenon continues if we increase the number of settings for the observers to three and four. The investigations were also extended to higher dimensional systems, for which we tested a few specially chosen states. The phenomenon seems to persist, and is even more pronounced. As a by-product of our investigations we have found for two qutrits a relatively simple set of unitary transformations (defined by just three phases), which gives a map of violations of local realism which diverges from the one for the full U(3) group defined measurements by maximally just 1.5\%, and for a broad class of states leads to the same level of non-classicality. We have also studied some interesting mixed states, which contain bound entanglement.

\section{Description of the method} 

\begin{figure}[!ht]
\includegraphics[width=0.35\textwidth]{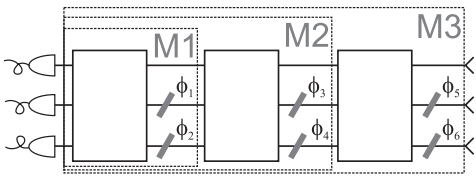}
\caption{\label{Ufig-jg} Measurement devices represented by unitary transformations M1, M2 and M3. The inputs are on the right hand side. Boxes represent unbiased symmetric three-port beamsplitters \cite{CONTROL,multiport}.}
\end{figure}

Let us move on to the quantum processes that we analyze, and the numerical approach. We begin with the two-qutrit case.
In our numerical analysis (called {\ttfamily steam-roller}) we consider a class of pure states of two qutrits, which contains all possible Schmidt decompositions:
\begin{equation}
|\psi(\alpha, \beta)\rangle = \cos \alpha |00\rangle + \sin \alpha (\cos \beta  |11\rangle +  \sin \beta |22\rangle).
\label{state-jg}
\end{equation}
Two spatially separated observers perform measurements of $m$ alternative local noncommuting trichotomic observables: $A_1, A_2, ... A_m$ for Alice and $B_1, B_2, ... B_m$ for Bob. We assume that they measure observables defined by a set of phase-shifters and (see Fig. \ref{Ufig-jg}):
\begin{itemize}
\item (M1) one unbiased three-port beam-splitter (tritter, for its properties, see\cite{multiport}),
\item (M2) two unbiased three-port beam-splitters,
\item (M3) three unbiased three-port beam-splitters,
\end{itemize}
and in the ultimate case
\begin{itemize}
\item (U(3)) any three-dimensional unitary transformation belongs to the U(3) group (for parametrization see e.g. \cite{SUN}).
\end{itemize}
A $d$ input and output unbiased multiport performs a unitary transformation described by the Fourier matrix $U_{kl}={d}^{-{1}/{2}}e^{i{(2kl\pi}/{d})}, $ where $k,l=0,1...,d-1.$ 
The unitary transformations M1 and M2, obviously cannot reproduce the full U(3) group. We have checked that this is also the case for M3 transformations. 

By saying that an experiment is local realistic we understand that it has a local realistic model for the assumed set of settings. 
In order to obtain the value of the critical visibility $v_{crit}$ to allow such models for any observables from a given class,
we follow the procedure of the kind first used in Refs. \cite{GRUCA, KASZLIKOWSKI} (but with a much more advanced implementation and structure). The method is described in the Appendix.
%%%%%%%%%%%%%%%%%%%%%%%%%%%%%%%%%%%%%%%%%%%%%%%%%%%%%%%%%%%%%%%%%%%%%%%%%%%%%%%%%%%%%%%

%%%%%%%%%%%%%%%%%%%%%%%%%%%%%%%%%%%%%%%%%%%%%%%%%%%%%%%%%%%%%%%%%%%%%%%%%%%%%%%%%%%%%%%%%

\section{Results} 

\begin{figure}[t]
\includegraphics[width=0.45\textwidth]{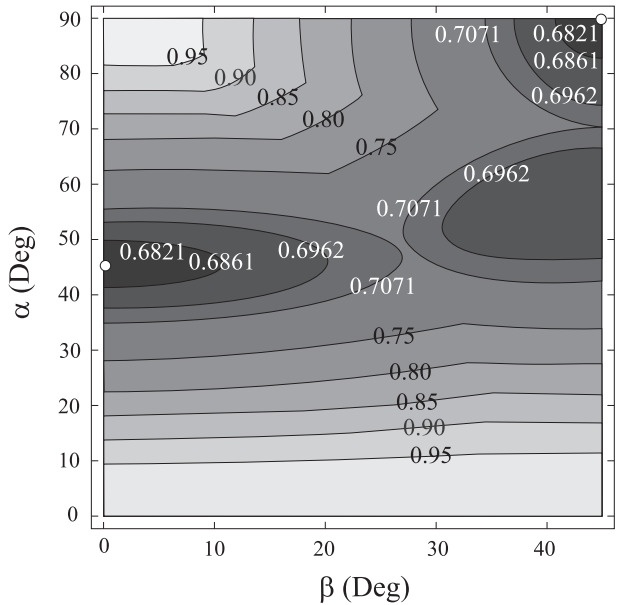}
\caption{\label{violmap} The map of critical visibilities $v_{crit}$, that is noise resistance of non-classical correlations for the state $|\psi(\alpha, \beta)\rangle = \cos \alpha |00\rangle + \sin \alpha (\cos \beta  |11\rangle +  \sin \beta |22\rangle)$ for the case in which both observers use any observables. If $v>v_{crit}$, there does not exist any local realistic model describing quantum probabilities of experimental events.}
\end{figure}

We have computed a map of strength of violation of local realism by pure two qutrit states. There one can find a range of the $\alpha$ and $\beta$ parameters of the state (\ref{state-jg}), for which the critical visibility is below the lowest known critical visibility ($0.6861$) in the case of the CGLMP inequality, for the asymmetric state $|\psi_{asym}\rangle$ found in ref. \cite{ACIN} (see Fig. \ref{violmap}).  The lowest critical visibility is 0.6821 and surprisingly corresponds to the Schmidt rank-2 states $|\psi_{sym}^{rank-2}\rangle$: $|\psi(90^{\circ}, 45^{\circ})\rangle = (|11\rangle + |22\rangle)/\sqrt{2}$, or $|\psi(45^{\circ}, 0^{\circ})\rangle = (|00\rangle + |11\rangle)/\sqrt{2}$ and $|\psi(45^{\circ}, 90^{\circ})\rangle = (|00\rangle + |22\rangle)/\sqrt{2}$ (outside the map, but trivially related with the previous one). We can observe this phenomenon by using the most general local three-dimensional observables (U(3)) and transformations M3, while for less general observables (e.g. M1, M2), it is unobservable. 

Despite the fact that the model noise well describes e.g. dark counts, one may criticize ``white'' noise resistance as a measure of non-classicality, see e.g. \cite{ADGL}. Thus, we have studied also other types of noise.
For a product noise admixture, of the form $\rho_A \otimes \rho_B$, where $\rho_{A(B)}$ is a reduced density matrix of the state (\ref{state-jg}) of the system A(B), the critical visibility for the state $|\psi_{sym}^{rank-2}\rangle$ is  equal to 0.7071, whereas for the states $|\psi_{asym}\rangle$, $|\psi_{sym}\rangle$ is the same as in the case of white noise admixture, respectively 0.6962, 0.6861. For a dephasing noise, $\cos^2{\alpha} |00\rangle\langle00| + \sin^2{\alpha}( \cos^2{\beta} |11\rangle\langle11| +  \sin^2{\beta} |22\rangle\langle22|)$, we observe violations of local realism for any state $|\psi(\alpha,\beta)\rangle$ ($\alpha>0$) and any amount of the noise ($v<1$). Such a feature is well known for qubits, however for qutrits thus far it was only a conjecture.

\begin{table*}
\begin{tabular}{c c c c c c c c} \hline \hline
$d$ & $|\psi_{sym}^{rank-d}\rangle$ & $|\psi_{asym}^{rank-d}\rangle$   & 
      $|\psi_{sym}^{rank-4}\rangle$ & $|\psi_{asym}^{rank-4}\rangle$   &
      $|\psi_{sym}^{rank-3}\rangle$ & $|\psi_{asym}^{rank-3}\rangle$   &
      $|\psi_{sym}^{rank-2}\rangle$   \\ \hline
3 & 0.6962 & 0.6861 & -  & - & - & - & 0.6821 \\
4 & 0.6906 & 0.6728 & -  & - & 0.6824  & 0.6725  & 0.6442 \\
5 & 0.6871 & 0.6632 & 0.6819 & 0.6637  & 0.6584  & 0.6485& 0.6071 \\
\hline \hline
\end{tabular}
\caption{\label{min} The critical visibilities for special states. 
The symbol  $|\psi_{sym}^{rank-k}\rangle$ for $k=d$ stands for a maximally entangled  state of two qu$d$its, that is with symmetric (all equal) Schmidt coefficients. If $k<d$ it is a state of two qu$d$its resembling a maximally entangled symmetric state for $k$-dimensional systems.  The states $|\psi_{asym}^{rank-k}\rangle$ represent in the case $k=d$ the one which maximally violates the CGLMP inequality for qu$d$its, whereas for $k<d$, they are  two qu$d$it states resembling the one optimal for the CGLMP inequality for a $k$ dimensional problem. The observers perform unrestricted measurements determined locally by U($d$).}
\end{table*}

We can also observe such phenomena in the 4-dimensional case. For the most general local 4-dimensional observables defined by the U(4) group, the critical visibility obtained for a symmetric Schmidt rank-2 state (e.g. $(|00\rangle + |11\rangle)/\sqrt{2}$) is equal to $0.6442$, whereas the critical visibility necessary for violation the CGLMP inequality is equal to $0.6728$ and $0.6906$ for the asymmetric \cite{CHEN} and symmetric \cite{CGLMP} states respectively (see Tab. \ref{min}).

The critical values for rank-2 states, Tab. \ref{min}, agree with those reported in \cite{ADGL}, where such numbers were found employing an {\em ad hoc} approach, aimed at showing conceptual problems related with the noise resistance as measure of non-classicality. Here we establish that these are the lowest possible critical visibilities for two-qutrit states and give a strong numerically supported conjecture that it is so for higher-dimensional systems. One could claim that a rank-2 state is effectively a two qubit state. However, as it is clear from our graphs, in the case of qutrits there is a continuous family of (strongly asymmetric) rank-3 states, which violate local realism stronger (in terms of noise resistance) than the optimal state for violation of the CGLMP inequality.

\begin{figure}[!h]
\includegraphics[width=0.40\textwidth]{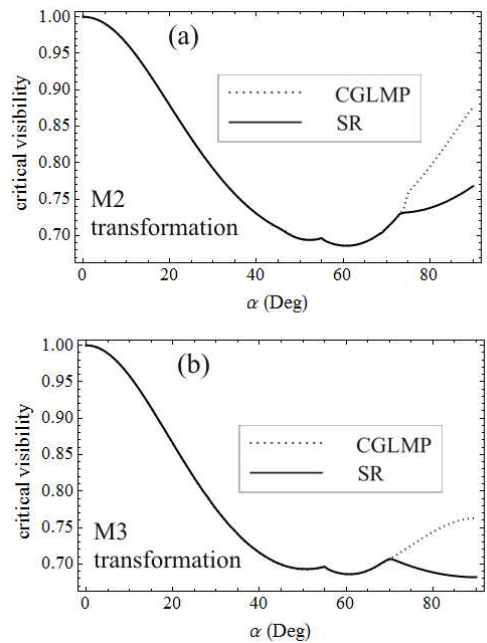}
\caption{\label{CGLMP-fig} The dotted lines represent the critical visibility, which is necessary to violate the CGLMP inequality by  states $|\psi(\alpha, 45^{\circ})\rangle$. The solid lines correspond to the critical visibility obtained with {\ttfamily steam-roller}. M2 (left graph) and M3 (right graph) transformations are considered here. The points at the curves with an undefined tangent indicate that different Bell inequalities are maximally violated by states to the left and right of the point.}
\end{figure}

We compared  our  computer code, which is  equivalent to a full set of Bell inequalities for the given problem, with the CGLMP inequality. For two qutrit states of the form $|\psi(\alpha, 45^{\circ})\rangle$,
if we restrict observables to  M1 type, the numerical method gives the same\footnote{All comparisons are accurate to four decimal places.} critical visibility as the CGLMP inequality \cite{CGLMP}. Note that such observables were used in earlier works, refs \cite{KASZLIKOWSKI,CGLMP,ADGL,ACIN,CHEN,ZOHREN}. 
 In the case of observables of M2 kind, for $\alpha > 73^{\circ}$ the critical visibility obtained with {\ttfamily steam-roller} is lower than predicted by CGLMP inequality (see Fig. \ref{CGLMP-fig}a). For observables parametrized by M3 (and full U(3)) the advantage of the numerical method is even better visible and occurs already for $\alpha > 70^{\circ}$ (see Fig. \ref{CGLMP-fig}b). 

\begin{figure}[!ht]
\includegraphics[width=0.45\textwidth]{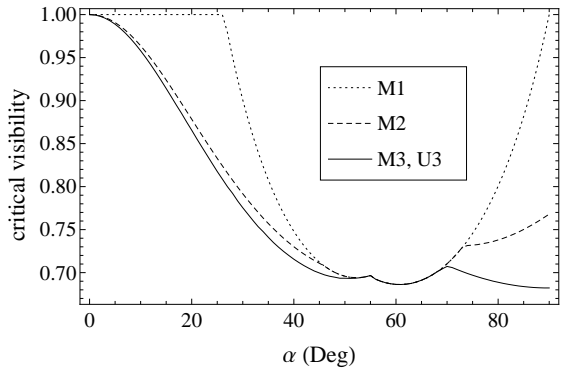}
\caption{\label{2Dplot} The critical visibilities for  states $|\psi(\alpha, 45^{\circ})\rangle$. Each line corresponds to a different type of the unitary transformations (M1, M2, and M3). Up to four settings per observer were used, the relations did not change.}
\end{figure}

For the two-qutrit case, there are some ranges of the parameter $\alpha$ in $|\psi(\alpha, 45^{\circ})\rangle$, for which: 
\begin{itemize}
\item the observable generated by M2 gives a better critical visibility than the ones obtained for the observable M1 ($\alpha <  49^{\circ}$ and $\alpha > 73^{\circ}$). 
\item the observable for M3 gives a better critical visibility than the ones obtained for the observable M2 ($\alpha <  54^{\circ}$ and $\alpha > 70^{\circ}$). 
\end{itemize} 
All of these cases are presented in Fig. \ref{2Dplot}. The critical visibilities obtained for the M3 and U(3) observables are the same for the $|\psi(\alpha, 45^{\circ})\rangle$ state. For other values of $\beta$ the highest difference $\delta v_{crit}^{\rm U(3)-M3}(\%)$ between critical visibilities obtained for these observables is less than 1\%.  
If one increases the number of settings, up to four on each side, this does not reduce the critical visibility for any of the types of observables.

\subsection{Almost perfect two-qutrit Bell interferometer}
\begin{figure}[t]
\includegraphics[width=0.40\textwidth]{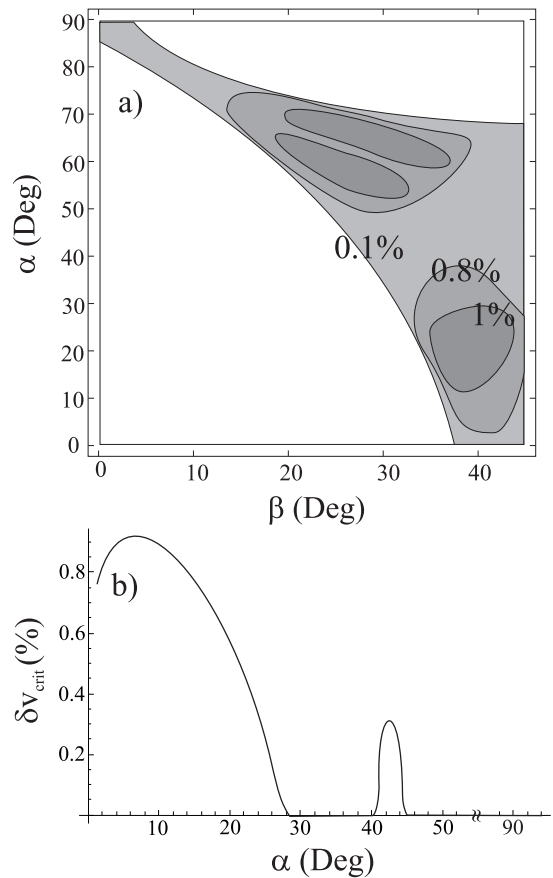}
\caption{\label{violmapprec} a) Differences between critical visibilities obtained for the U(3) and M3 (with $\phi_1=\phi_5=\phi_6=0$) transformations for the state $|\psi(\alpha, \beta)\rangle$. Peak points are lower than 1.5\%; b) Differences between critical visibilities obtained for the U(3) and M3 (with $\phi_1=\phi_5=\phi_6=0$) transformations for the states $|\psi(\alpha, 45^{\circ})\rangle$.}
\end{figure}

Let us consider the M3 interferometer shown in Fig. \ref{Ufig-jg} with $\phi_1=\phi_5=\phi_6=0$. The differences between critical visibilities obtained for such transformation and U(3) are shown in Fig. \ref{violmapprec}a for the state $|\psi(\alpha, \beta)\rangle$ and in Fig. \ref{violmapprec}b for the state $|\psi(\alpha, 45^{\circ})\rangle$. They are less than 1.5\% for any $\alpha$ and $\beta$, and in many cases the same (including specific points related to the symmetric $|\psi_{sym}\rangle$ and asymmtric $|\psi_{asym}\rangle$ state). Thus if one is interested in situations in which violations of local realism play an essential role this relatively simple device might be optimal.

\subsection{Question of violation of local realism by bound entangled states} 
As we have at our disposal a tool which allows to test violation of local realism for any local observables, we also analyzed some classes of two-particle bound entangled states, namely:
the three-dimensional Bennett state \cite{BENNETT}, the three-dimensional Horodecki state \cite{HORODECKIBOUND} and its generalization to the four- and five-dimensional states \cite{CHRUSCINSKI}, the four-dimensional Pankowski -- Horodecki state \cite{PANKOWSKI}. We established that for U(3), and in the last case U(4) transformations (that is all possible local measurements with two settings per observer) one does not observe violations of local realism.

\section{Final remarks} 
We have charted unknown territories in the map of non-classicality of two qutrit states, and made some excursions to higher dimensional problems. The obtained results were difficult to anticipate. Our method is quite flexible and we plan to move on to the yet uncharted multi qudit states. Note, that the computer code, after special processing of experimental data is able to decide whether the data allow a local realistic model or not. Such applications will be found in forthcoming manuscripts. The general theory of the code itself, and particular solutions, will be presented in ref. \cite{GONDZIO}. 

{\em Acknowledgments.---}  The work is supported by EU program QESSENCE
(Contract No.248095) and  MNiSW (NCN) Grant no. N202 208538. 
WL acknowledges financial support from European Social
Fund as a part of the project ``Educators for the elite
- integrated training program for PhD students, postdocs
and professors as academic teachers at University of
Gda\'nsk'' within the framework of Human Capital Operational
Programme, Action 4.1.1, Improving the quality
of educational offer of tertiary education institutions. JG akcnowledges DS IFTiA UG.

\appendix*

\section{Description of the method}

\maketitle

The critical visibility is calculated with the use of the simplex algorithm \cite{simplex} from
the GLPK library \cite{GLPK}. This critical visibility is dependent on the angles 
(parameters of the observables) and it can be shown that it is in
fact a continuous but non-differentiable function, of the angles \cite{grucamgr}. 
An intuitive explanation of why this is the case will be given in the last paragraph. 
The critical visibility function can be minimized with
the use of non-linear optimization algorithms. Since it is
non-differentiable we are limited to the ones which do not require
the derivative of the optimized function. We use the established
Nelder Mead method \cite{dsm}. Despite problems raised in \cite{cytJg1} where it is shown that
the method does not converge to a stationary point in the general
case, in our case it converges nicely when applied to the critical
visibility function. This was tested
extensively by comparison with both results derived analytically and known from
other sources. It should be noted, however, that while the
search for critical visibility is guaranteed to return the correct
result \cite{simplex}, the Nelder-Mead method may return a local minimum.
Therefore, one can be sure that the actual minimal critical visibility
is not greater than the one given here (but it might be smaller). 

For qutrit experiments  with two alternative measurement settings per observer ($m=2$), realistic models are equivalent to the existence of a joint probability distribution $p_{lr}(a_1,a_2,b_1,b_2)$, where $a_i = {0,1,2} ~(i=1,2)$ denotes the result of the measurement of Alice's $i$-th observable (Bob's results are denoted by $b_k = {0, 1, 2} ~(k=1,2)$). Quantum predictions for the probabilities should be given, if the model exists,  by marginal sums
\begin{equation}
P(a_i, b_k| A_i, B_k) =  \sum_{a_{i'}, b_{k'} = 0}^2 p_{lr}(a_1, a_2, b_1, b_2),\label{marginals}
\end{equation}
where $P(a_i, b_k| A_i, B_k)$ denotes the probability of obtaining the result $a_i$ by Alice and $b_k$ by Bob, if they measure the observables $A_i$ and $B_k$, respectively and $i' \neq i, k' \neq k$.  

If we admix some amount of white noise to the two qutrit state, we obtain the quantum probabilities
$P^{v}_{QM}(a_i, b_k| A_i, B_k) \equiv P(a_i, b_k| A_i, B_k)$ in the form of 
\begin{equation}
P(a_i, b_k | A_i, B_k) = v P_{QM}(a_i, b_k | A_i, B_k) + \frac{1}{3^2}(1-v),
\label{probability}
\end{equation}
where $P_{QM}$ denotes the quantum-mechanically calculated probability for the state without the noise admixture.

For $v=1$, it is known that for some entangled states, no local realistic probability distribution $p_{lr}(a_1, a_2, b_1, b_2)$ which could satisfy the set of equalities (\ref{marginals}) exists. Therefore, no local realistic model can reproduce the predictions of quantum mechanics for these entangled states. However, for such states there always exists the critical visibility $v_{crit}$ such that for $v\leqslant v_{crit}$ there exists a local realistic probability distribution $p_{lr}(a_1, a_2, b_1, b_2)$ that satisfies the set of equalities (\ref{marginals}). Our task --- for a given state $\rho$ --- is to find settings for which the critical visibility is minimal. 

The computation of $v_{crit}$ (for the given parameters) is equivalent to finding the set of $3^4$ positive numbers (the values of $p_{lr}(a_1,a_2,b_1,b_2)$) summing up to $1.0$ and fulfilling the $2^2 \cdot 3^2 = 36$ conditions given by (\ref{marginals}) such that $v$ is maximal. Therefore, $v$ and $p_{lr}(a_1,a_2,b_1,b_2)$ can be treated as variables lying in a $(3^4 + 1)$-dimensional real space. The set of linear conditions (\ref{marginals}) and the conditions $0 \leq v \leq 1; 0 \leq p_i \leq 1$ define a convex set in this space. Our task is to find a vertex of the set with the largest $v$ coordinate. This is done by means of linear programming \cite{GRUCA, KASZLIKOWSKI}. Of course one can easily formulate generalization of the presented approach to more settings per observer and systems of higher dimensionality. However in such cases the computing time increases, and thus we studied only specific interesting examples. 

The 36 probabilities $P(a_i, b_k|A_i,B_k)$ define a point $S$ in a
36-dimensional space, which might be inside, on the boundary or
outside the Pitovsky-Bell polytope in that space \cite{PIT}. The point $S$ in this space is within
(i.e. inside or on the boundary of) the Pitovsky-Bell polytope if and only if
the set of equalities (\ref{marginals}) is  satisfied by the coordinates of $S$. Let us
consider a state $|S\rangle$ and a set of observables $A_i, B_k$ for which the
corresponding point $S$ is outside the Bell polytope (for $v=1$). The position of $S$ is dependent on $v$, with $S(v=0)$ well inside the
polytope and $S(v=v_{crit})$ on a facet of the polytope. Given
(\ref{probability}) we know that $S(v)$ is a linear function in the 36-dimensional
space. The critical visibility function for slightly perturbed angles
(parameters of the observables) can now be calculated by recalculating
all $P(a_i,b_k|A_i,B_k)$ probabilities and connecting $S(v=0)$ and the new
$S'(v=1)$ with a straight line. $S(v=0)$ will not change its position since
for visibility equal to 0 the observables do not matter. $S'(v=1)$ will
be slightly perturbed. Thus the new $S'(v=v_{crit})$ point will also be
``just next to'' the previous critical visibility point and hence --- the
critical visibility function is continuous. It is also differentiable
as long as we stay on the same facet of the Bell polytope. But if we
perturb the angles in a vicinity of an edge of the Bell polytope,
differentiability is lost. Hence, as mentioned before, the critical
visibility is a continuous but non-differentiable function of parameters defining the observables.


\begin{thebibliography}{99}
\bibitem{BELL} J. S. Bell, Physics {\bf 1}, 195 (1964).
\bibitem{CLAUSER} J.F. Clauser and A. Shimony, Rep. Prog. Phys.
{\bf 41}, 1881 (1978)
\bibitem{MERMIN} N. D. Mermin, Phys. Rev. D 22 356 (1980); N. D. Mermin and G. M. Schwarz, Found. Phys. {\bf 12}, 101 (1982); M.
Ardehali, Phys. Rev. D {\bf 44}, 3336 (1991); A. Garg and N.
D. Mermin, Phys. Rev. Lett. {\bf 49}, 901 (1982); similar results were obtained by K. W\'odkiewicz, Acta. Phys. Pol.
A {\bf 86}, 223 (1994), who considered local projections on spin-coherent states.
\bibitem{GHZ}  D.M. Greenberger, M.A. Horne and A. Zeilinger, in Bell's theorem and the Conception of the Universe, edited by M.
Kafatos (Kluwer Academic, Dordrecht, 1989).
\bibitem{PERES} A. Peres, Phys. Rev. A {\bf 46}, 4413 (1992), N. Gisin and A.
Peres, Phys. Lett. A {\bf 162}, 15-17 (1992).
\bibitem{KLYSHKO} D.N. Klyshko, Phys. Lett. A {\bf 132}, 299 (1988).
\bibitem{CONTROL} A. Zeilinger, H.J. Bernstein, D.M. Greenberger, M.A.
Horne, and M. \.Zukowski, in  Quantum Control and Measurement, eds. H. Ezawa and Y. Murayama (Elsevier,
1993); A. Zeilinger, M. \.Zukowski, M.A. Horne, H.J. Bernstein and D.M. Greenberger, in Quantum Interferometry,
eds. F. DeMartini, A. Zeilinger, (World Scientific, Singapore, 1994).
\bibitem{multiport}  M. \.Zukowski, A. Zeilinger, and M. A. Horne, Phys. Rev. A {\bf 55}, 2564 (1997). Unbiased multiports were earlier
called symmetric ones.
\bibitem{RECK} M. Reck, A. Zeilinger, H. J. Bernstein, and P. Bertani, Phys. Rev. Lett. {\bf 73}, 58 (1994).
\bibitem{KASZLIKOWSKI} D. Kaszlikowski, P. Gnaci\'nski, Marek \.Zukowski, W. Miklaszewski, and A. Zeilinger, Phys. Rev. Lett. {\bf 85}, 4418 (2000).
\bibitem{CGLMP} D. Collins, N. Gisin, N. Linden, S. Massar, and S. Popescu, Phys. Rev. Lett. {\bf 88}, 040404 (2002).
\bibitem{ADGL} A. Ac\'in, T. Durt, N. Gisin, J. I. Latorre, Phys. Rev. A {\bf 65}, 052325 (2002).
\bibitem{ACIN} A. Ac\'in, R. Gill, and N. Gisin, Phys. Rev. Lett. {\bf 95}, 210402 (2005).
\bibitem{CHEN} J.-L. Chen, Ch. Wu, L.C. Kwek, C.H. Oh, and M.-L. Ge, Phys. Rev. A {\bf 74}, 032106 (2006).
\bibitem{ZOHREN} S. Zohren and R. D. Gill, Phys. Rev. Lett. {\bf 100}, 120406 (2008).
%\bibitem{simplex} G. B. Dantzig, Linear Programming and Extensions, Princeton University Press, Princeton, NJ (1963).
%\bibitem{GLPK} GNU Linear Programming Kit, Version 4.31, http://www.gnu.org/software/glpk/
%\bibitem{dsm} J. A. Nelder, R. Mead, Computer Journal {\bf 7}, 308 (1965).
%\bibitem{cytJg1} K. I. M. Mckinnon, SIAM J. Opt {\bf 9} 148, (1998).
%\bibitem{PIT} I. Pitovsky, Quantum Probability-Quantum Logic, Springer, Berlin (1989).
%\bibitem{scipy} SciPy Python scientific computing package, Version 0.7.0, http://www.scipy.org/
\bibitem{GRUCA} J. Gruca, W. Laskowski, M. \.Zukowski, N. Kiesel, W. Wieczorek, C. Schmid and H. Weinfurter, Phys. Rev. A {\bf 82}, 012118 (2010).
%\bibitem{grucamgr} J. Gruca, Nonclassical properties of multiqubit quantum states applicable in quantum information science, Master's thesis, University of Gdansk, 2008.
\bibitem{SUN} T. Tilma and E. C. G. Sudarshan, J. Phys. A: Math. Gen. {\bf 35}, 10467 (2002).
\bibitem{BENNETT} C.H. Bennett, D.P. DiVincenzo, T. Mor, P.W. Shor, J.A. Smolin, B.M. Terhal, Phys.Rev.Lett. {\bf 82} 5385 (1999).
\bibitem{HORODECKIBOUND} P. Horodecki, Phys. Lett. A {\bf 232}, 333 (1997).
\bibitem{CHRUSCINSKI} D. Chru\'sci\'nski, A. Rutkowski, Phys. Lett. A {\bf 375}, 2793 (2011).
\bibitem{PANKOWSKI} \L. Pankowski, M. Horodecki, IEEE Trans on Inf Theor, {\bf 54}, 2621 (2011).
\bibitem{GONDZIO} J. Gondzio, J. Gruca, J.A.J. Hall, W. Laskowski, M. \.Zukowski, in preparation.
\bibitem{simplex}
G. B. Dantzig, Linear Programming and Extensions, Princeton University Press, Princeton, NJ (1963).
\bibitem{GLPK}
GNU Linear Programming Kit, Version 4.31, http://www.gnu.org/software/glpk/
\bibitem{dsm}
J. A. Nelder, R. Mead, Computer Journal {\bf 7}, 308 (1965).
\bibitem{cytJg1} K. I. M. Mckinnon, SIAM J. Opt {\bf 9} 148, (1998).
\bibitem{PIT} I. Pitovsky, Quantum Probability-Quantum Logic, Springer, Berlin (1989).
\bibitem{scipy} SciPy Python scientific computing package, Version 0.7.0, http://www.scipy.org/
\bibitem{grucamgr} J. Gruca, Nonclassical properties of multiqubit quantum states applicable in quantum information science, Master's thesis, University of Gdansk, 2008.
\end{thebibliography}
\end{document}